\documentclass[sigconf]{acmart}

\usepackage[skip=0pt]{caption}
\usepackage{makecell}
\usepackage{etc}
\usepackage{float}
\usepackage{algorithm}
\usepackage{algpseudocode}
\usepackage{graphicx}
\usepackage{multirow}
\usepackage{textcomp}
\usepackage{booktabs}
\usepackage{subcaption}
\usepackage[export]{adjustbox}
\usepackage{balance}
\usepackage{csquotes}
\usepackage{textcomp}
\usepackage{arydshln}
\usepackage{multirow}

\PassOptionsToPackage{colorlinks,bookmarksopen,bookmarksnumbered,citecolor=red, urlcolor=black}{hyperref}

\definecolor{commoncolor}{HTML}{000000}       
\definecolor{encoderonlycolor}{HTML}{00529B}   
\definecolor{encoderdecodercolor}{HTML}{C46210} 
\definecolor{mergecolor}{HTML}{4F7942}

\newcommand{\rqone}{To what extent do merging strategies on language models for code save computational resources?}
\newcommand{\rqtwo}{What is the impact of merging strategies on the performance of language models of code on downstream SE tasks?}
\newcommand{\rqthree}{What is the impact of the position of the merging layer within the language model of code?}
\newcommand{\rqfour}{Which merging configuration (\ie{} strategy and layer position) yield the best trade-off between performance and computational requirements?}
\DeclareTextFontCommand{\emp}{\bfseries}

\newcommand{\common}[1]{\textcolor{commoncolor}{#1}}
\newcommand{\encoderonly}[1]{\textcolor{encoderonlycolor}{#1}}
\newcommand{\decoderonly}[1]{\textcolor{encoderdecodercolor}{#1}}
\newcommand{\merge}[1]{\textcolor{mergecolor}{#1}}

\AtBeginDocument{%
  }

\setcopyright{acmlicensed}
\copyrightyear{2026}
\acmYear{2026}
\acmDOI{XXXXXXX.XXXXXXX}

\acmConference[Conference acronym 'XX]{Make sure to enter the correct
  conference title from your rights confirmation email}{June 03--05,
  2018}{Woodstock, NY}

\keywords{Transformers, Merging, Efficient Fine-tuning}

\ccsdesc[500]{Software and its engineering~Software libraries and repositories}
\ccsdesc[300]{Computing methodologies~Neural networks}
\ccsdesc[500]{Software and its engineering~Software performance}
\ccsdesc[500]{Software and its engineering~Correctness}

\begin{document}
\raggedbottom

\title{On the Effect of Token Merging on Pre-trained Models for Code}


\author{Mootez Saad}
\affiliation{%
  \institution{Dalhouise University}
  \city{Halifax}
  \country{Canada}}
\email{mootez@dal.ca}

\author{Hao Li}
\affiliation{%
  \institution{Queen’s University}
  \city{Kingston}
  \country{Canada}}
\email{hao.li@queensu.ca}

\author{Tushar Sharma}
\affiliation{%
  \institution{Dalhouise University}
  \city{Halifax}
  \country{Canada}}
\email{tushar@dal.ca}

\author{Ahmed E. Hassan}
\affiliation{%
  \institution{Queen’s University}
  \city{Kingston}
  \country{Canada}}
\email{ahmed@cs.queensu.ca}
\begin{abstract}
Tokenization is a fundamental component of language models for code. It involves breaking down the input into units that are later passed to the language model stack to learn high-dimensional representations used in various contexts, from classification to generation. However, the output of these tokenizers is often longer than that traditionally used in compilers and interpreters. This could result in undesirable effects, such as increased computational overhead. In this work, we investigate the effect of merging the hidden representations of subtokens that belong to the same semantic unit, such as subtokens that form a single identifier. We propose two strategies: one based on averaging the representations and another that leverages a learning-based approach. Both methods can be seamlessly integrated with existing language models for code. We conduct experiments using six language models for code: \cb{}, \gcb{}, \unx{}, \ctf{}, \ctfpttz{}, and \ctfpssz{}, across three software engineering tasks: \textit{vulnerability detection}, \textit{code classification}, and \textit{code translation}. Results show that these strategies can reduce the number of floating-point operations by $1\%$ to $19\%$. Regarding downstream performance, the most significant degradation was observed in the \textit{vulnerability detection} task, where the F1 score decreased by $1.82$ points compared to the baseline. In contrast, for \textit{code translation}, we observed an improvement of $2.47$ points in CodeBLEU. This work contributes to the broader effort of improving language models for code across multiple dimensions, including both computational efficiency and downstream performance.
\end{abstract}

\maketitle

\section{Introduction}\label{sec:introduction}

Pre-trained language models for code have emerged as powerful tools for various software engineering~(SE) tasks such as vulnerability detection and code search. A key step for applying these models to source code is \textit{tokenization}, which converts code into a sequence of smaller units called \textit{tokens} that the model can process~\cite{karampatsis_bigcode_2020, feng2024toma}. However, tokenizing code presents unique challenges such as accurately handling nested statements, maintaining precise lexeme boundaries, and capturing the structural and syntactic characteristics of programming languages~(PLs) that differ markedly from natural languages~(NLs)~\cite{shen2022codesyntax}.

\begin{figure}[t]
    \centering
    \begin{subfigure}[b]{0.53\columnwidth}
        \centering
        \includegraphics[width=\textwidth]{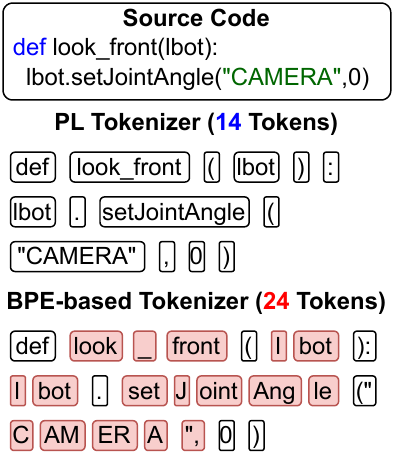}
        \caption{Tokenization example}
        \label{fig:identifiers_token}
    \end{subfigure}
    \hfill
    \begin{subfigure}[b]{0.45\columnwidth}
        \centering
        \includegraphics[width=\textwidth]{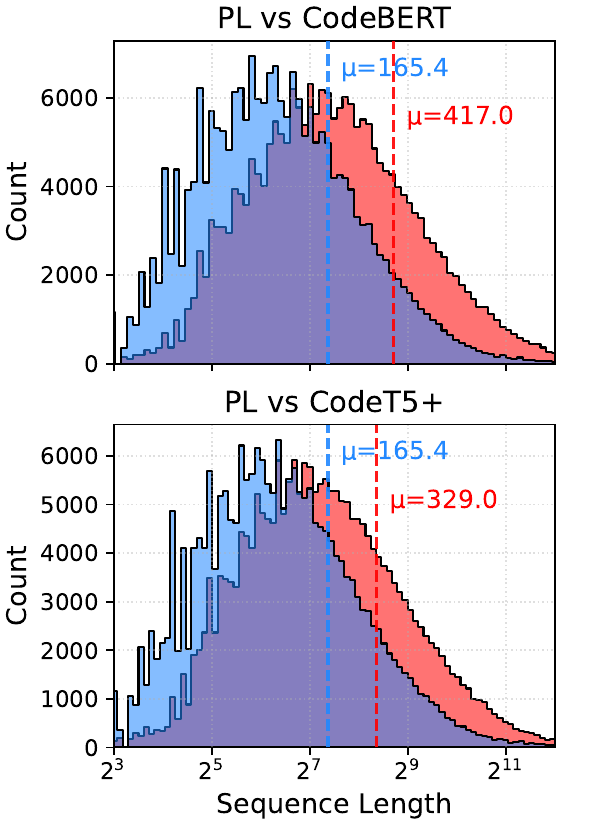}
        \caption{PL vs. BPE tokenizers}
        \label{fig:token_length_compare}
    \end{subfigure}
    \caption{Left: A motivating example showing how a Python code snippet is tokenized using a PL-based tokenizer implemented using tree-sitter compared to a BPE-based tokenizer. Right: Distributions of sequence lengths produced by \cb{} and {\sc CodeT5+} tokenizers, and a PL-based tokenizer.}
    \label{fig:python}
\end{figure}

Many models use byte-pair encoding~(BPE)~\cite{gage1994bpe, sennrich2016bpe}, a subword-based tokenization method originally designed for natural language processing~(NLP) to handle out-of-vocabulary~(OOV) words. BPE has also been applied to source code, helping address OOV issues~\cite{karampatsis_bigcode_2020}. However, around $70$\% of a software system's source code consists of identifiers~\cite{deissenboeck_concise_2006}, which developers can create freely, often leading to complex and compounded identifier names~\cite{karampatsis_bigcode_2020}. BPE tends to split these identifiers into multiple subword tokens, increasing sequence length compared to grammar-based PL tokenizers like tree-sitter~\cite{tree-sitter}.

For instance, as shown in Figure~\ref{fig:identifiers_token}, the BPE-based tokenizer (\cb{}) breaks identifiers into smaller subwords, while a PL tokenizer (tree-sitter) preserves them as single units. Similarly, Figure~\ref{fig:token_length_compare} compares sequence length distributions for C/C++ code samples from the \bigvul{}~\citep{fan2020bigvul} dataset, showing that BPE-based tokenizers produce sequences around $2.0$ times longer on average than the PL tokenizer. Longer sequences increase computational costs and may reduce the model's ability to properly capture code semantics.

Rather than designing a new tokenization method, this paper investigates post-tokenization strategies to merge tokens within the model itself. We explore various token merging strategies that aim to reduce sequence length without sacrificing performance in code-related tasks. These strategies aim to address inefficiencies in tokenized code sequences and provide more compact representations that preserve semantic information. By reducing sequence length, our methods have the potential to improve model performance on SE tasks, making these models both more effective and resource-efficient when applied to code. To achieve this, we address the following research questions~(RQs):


\begin{enumerate}
\item [\textbf{RQ1.}] \textbf{\rqone} 
We develop and implement multiple token merging strategies and apply them to SE datasets across different programming languages.
Intuitively, the less data a model processes, the more computation is saved. With this question, we aim to confirm and quantify these computational savings compared to the baseline models without merging strategies.

\item [\textbf{RQ2.}] \textbf{\rqtwo}
We fine-tune language models of code on various downstream SE tasks with and without the merging strategies. By measuring task performance with metrics such as F1 score and CodeBLEU score, we analyze how these merging strategies affect model performance.

\item [\textbf{RQ3.}] \textbf{\rqthree}
The merging strategies can be integrated at various positions within the language model of code. Through this research question, we investigate the effects of such design choices.

\item [\textbf{RQ4.}] \textbf{\rqfour}
As a continuation of the previous question, we aim to determine which configuration provides the best balance between model performance and computational savings.
\end{enumerate}

The main contributions of this paper are as follows.
\begin{itemize}
    \item We introduce and implement two token merging strategies that aim to address the limitation of language model tokenization and its side effects, without modifying it.
    \item We evaluate the impact of our token merging strategies on language model performance in downstream SE tasks, demonstrating their potential to enhance model efficiency and effectiveness.
    \item We conduct extensive experiments to investigate the effect of the position where merging occurs on the downstream performance and computational savings.
    \item We provide a replication package~\cite{replication}, which consists of the implementation of the proposed merging strategies, scripts for running the experiments, and the experiment results.
\end{itemize}

\textbf{Paper Organization.} The rest of this paper is organized as follows. Section~\ref{sec:background} provides background information. Section~\ref{sec:related_work} discusses related work. Section~\ref{sec:methodology} presents the merging strategies in our study. Section~\ref{sec:results} presents the results of our four RQs. Section~\ref{sec:implications} discusses the implications of our results. Section~\ref{sec:threats} discusses the threats to the validity of our study. Section~\ref{sec:conclusion} concludes this paper.
\section{Background}\label{sec:background}

In this section, we describe the mechanisms of tokenization and Transformer layers in language models.

\subsection{Tokenization in Language Models of Code}

Tokenization is a fundamental preprocessing step that involves converting raw text into a sequence of discrete units called tokens, which serve as the input to the model. Tokenization is crucial because it directly affects the model's ability to interpret and generate coherent text representations. The tokenization process for natural language typically aims to balance vocabulary size with the need for efficient sequence processing, often using subword-based tokenization methods that handle out-of-vocabulary words effectively.

One of the most commonly used tokenization methods in natural language processing~(NLP) is BPE~\cite{gage1994bpe, sennrich2016bpe}. BPE is a subword-based algorithm that iteratively merges pairs of frequently co-occurring characters or character sequences, allowing for a compact representation of vocabulary that includes both common and rare words. This approach helps to limit the vocabulary size while enabling the model to handle rare and compound words by breaking them down into smaller units, or subwords. BPE has demonstrated effectiveness in NLP applications by providing a balance between capturing semantic meaning and maintaining manageable sequence lengths.

However, when applied to code in programming languages (PLs), BPE encounters unique challenges. Unlike natural language text, code consists of structured syntax and long identifiers (e.g., \enquote{getUserDetails}, \enquote{fetchDataFromAPI}) that are not naturally segmented into common subwords. As a result, BPE often splits identifiers and other domain-specific tokens into multiple subword tokens. For instance, \enquote{getUserDetails} may be split into \enquote{get}, \enquote{User}, and \enquote{Details}, resulting in a longer token sequence than necessary. This can inflate the sequence length, increase computational costs, and may result in other unwanted side effects.

\subsection{The Transformer Architecture}

The Transformer architecture~\cite{vaswani2017attention} has become the backbone of modern language models of code, offering an efficient and scalable framework for processing long sequences. Transformers consist of stacked layers, each containing a multi-head self-attention mechanism (\mha{}) and position-wise feed-forward networks (\ffnn{}). The self-attention mechanism enables the model to capture dependencies between tokens at varying distances, allowing for contextual understanding of local and global patterns within a sequence. To elucidate the flow of data through these layers, we formalize the key operations below. Note that we confine this subsection to the encoder module of the Transformer model for simplicity.

Given an input sequence \(\mathbf{x} = (x_1, x_2, \dots, x_n)\), where \(n\) is the sequence length, each token \(x_i\) is mapped to a vector \(\mathbf{e}_i \in \mathbb{R}^{d_{\text{model}}}\) via an embedding matrix, producing \(E \in \mathbb{R}^{n \times d_{\text{model}}}\). Positional encodings \(\mathbf{p}_i \in \mathbb{R}^{d_{\text{model}}}\) are added to incorporate sequential order:
\[
E' = E + P, \quad \text{where} \quad P = [\mathbf{p}_1, \mathbf{p}_2, \dots, \mathbf{p}_n]^T,
\]
and \(E' \in \mathbb{R}^{n \times d_{\text{model}}}\) is passed to the first Transformer layer.

Each Transformer layer processes an input \(X \in \mathbb{R}^{n \times d_{\text{model}}}\) (with \(X = E'\) for the first layer) through two main sublayers. The first sublayer is the \mha{} layer, which computes attention scores across \(h\) heads to determine the influence of each token on every other token. For each head \(h\), the input is projected into:
\[
Q_h = X W_h^Q, \quad K_h = X W_h^K, \quad V_h = X W_h^V \in \mathbb{R}^{n \times d_k},
\]
where \(W_h^Q, W_h^K, W_h^V \in \mathbb{R}^{d_{\text{model}} \times d_k}\), and \(d_k = d_{\text{model}} / h\). The attention output for each head is:
\[
\text{head}_h = \text{softmax}\left( \frac{Q_h K_h^T}{\sqrt{d_k}} \right) V_h \in \mathbb{R}^{n \times d_k}.
\]
These are concatenated and linearly transformed:
\[
\text{MHA}(X) = \text{Concat}(\text{head}_1, \dots, \text{head}_h) W^O \in \mathbb{R}^{n \times d_{\text{model}}},
\]
with \(W^O \in \mathbb{R}^{d_{\text{model}} \times d_{\text{model}}}\). A residual connection and layer normalization are applied:
\[
X_{\text{MHA}} = \text{LayerNorm}(X + \text{MHA}(X)) \in \mathbb{R}^{n \times d_{\text{model}}}.
\]
This mechanism is particularly powerful for code, as it captures structural dependencies such as function calls or variable definitions spanning multiple lines.

Then, the \ffnn{} layer is applied independently to each token's vector in \(X_{\text{MHA}}\):
\[
\text{FFNN}(\mathbf{x}) = W_2 \cdot \text{ReLU}(W_1 \mathbf{x} + \mathbf{b}_1) + \mathbf{b}_2,
\]
where \(\mathbf{x} \in \mathbb{R}^{d_{\text{model}}}\), \(W_1 \in \mathbb{R}^{d_{\text{model}} \times d_{\text{ff}}}\), \(W_2 \in \mathbb{R}^{d_{\text{ff}} \times d_{\text{model}}}\), and \(d_{\text{ff}}\) is typically larger than \(d_{\text{model}}\). The output is:
\[
\text{FFNN}(X_{\text{MHA}}) \in \mathbb{R}^{n \times d_{\text{model}}},
\]
followed by another residual connection and layer normalization:
\[
X_{\text{out}} = \text{LayerNorm}(X_{\text{MHA}} + \text{FFNN}(X_{\text{MHA}})) \in \mathbb{R}^{n \times d_{\text{model}}},
\]
which becomes the input to the next layer. As inputs go through each layer, the model progressively refines its representation, capturing increasingly abstract features critical for understanding code syntax and semantics needed to solve various software engineering tasks.



\section{Related Work}\label{sec:related_work}

Tokenization plays an important role in adapting language models for SE tasks, yet existing tokenization methods introduce inefficiencies that can impact model performance. For instance, \citet{mastropaolo_studying_2021} propose a pre-tokenization abstraction approach that replaces domain-specific elements with placeholders (\eg{} \texttt{`VAR\_1'}, \texttt{`METHOD\_1'}) to create more compact inputs. While this reduces vocabulary size and improves performance, it requires modifying the vocabulary before training and inference. In contrast, our work preserves the original tokenization process by applying post-tokenization modifications without the need to modify the vocabulary, making it more flexible and compatible with pre-trained models.

Adaptive token reduction techniques have also been explored in both language and vision domains. \citet{casas_combining_2020} demonstrates that merging subword units into higher-level representations within Transformer encoders can yield efficient representations without compromising performance. Similarly, frameworks such as the Learned Thresholds Token Merging and Pruning for vision transformers~\citep{bonnaerens_ltmp_2023} show that dynamic, layer-wise token merging can substantially lower computational overhead. These methods offer conceptual tools that we adapt to address the unique challenges of code tokenization.

Other research has focused on efficient code representations. Studies on Coding-PTMs~\citep{zhao_codeptms_2024} and DataMUX~\citep{murahari_datamux_2022} underscores that compact input sequences are essential for balancing performance with resource efficiency. While these studies emphasize model selection and input multiplexing, our work complements them by showing that post-tokenization merging can reduce sequence length while preserving semantic content.

Additionally, alternative strategies such as embedding transfer and task-adaptive tokenization have been proposed to tackle tokenization challenges. \citet{liu_subword_2021} introduce an embedding transfer mechanism to bridge the gap between pretraining and fine-tuning caused by subword segmentation, and \citet{liu_taskadapt_2023} suggest dynamic vocabulary adaptation to reduce token fragmentation in specialized contexts.

While inspired by the works from language and vision domains mentioned above, this work adapts token merging strategies specifically to address the unique challenges of working with source code, such as the proliferation of complex and compounded identifiers created by developers. In addition, a key contribution of this paper is a detailed investigation into where the merging should occur that has not been explored in existing studies. We systematically experiment with applying the merging layer after the embedding layer and after various intermediate Transformer layers to analyze the impact on performance and computational savings.







\section{Methodology}\label{sec:methodology}

\begin{figure}[t]
    \centering
    \includegraphics[width=\columnwidth]{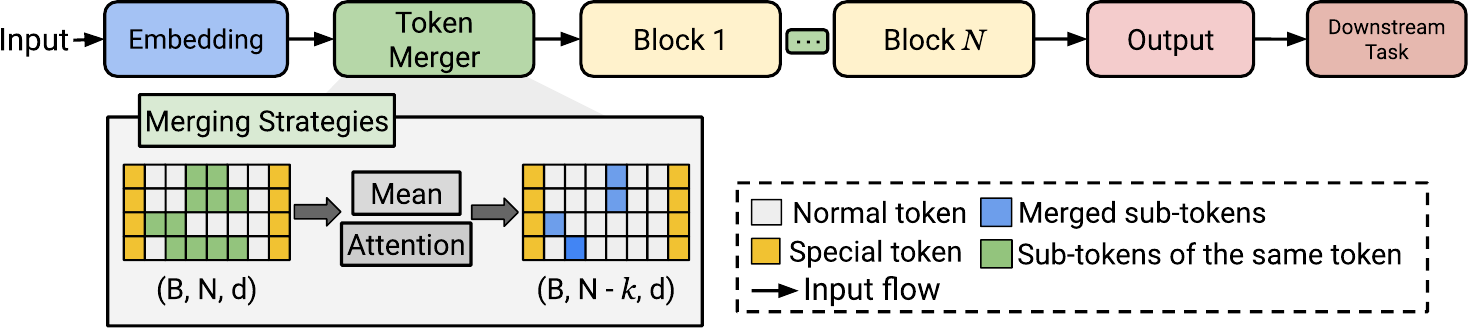}
    \caption{Overview of the proposed token merging strategy. We include input dimensions for better clarity. Specifically, $B$ refers to the batch size, $N$ is the number of tokens, and $d$ is the model's hidden dimension. After merging, the sequence length is reduced by $k$ tokens. }
    \label{fig:method_overview}
\end{figure}

In this section, we introduce our post-tokenization merging strategies and study subjects. Figure~\ref{fig:method_overview} provides an overview of our methodology.
\subsection{Token Merging Strategies}
\label{token_mering_strategies}

We explore two types of merging strategies: \textit{static} and \textit{learning-based}.
\subsubsection{Static Merging}
Static merging relies on a deterministic approach, aggregating subword representations using a fixed mathematical operation. Specifically, we use mean aggregation to merge subword tokens into a single representation by averaging their vector embeddings. Formally, let \(\mathbf{X} \in \mathbb{R}^{B \times N \times d}\) represent the sequence of token embeddings, where \(B\) is the batch size, \(N\) is the sequence length, and \(d\) is the hidden dimension. Tokens that correspond to the same semantic unit, such as an identifier, are grouped into a set \(S_i\). The merged representation \(\mathbf{X'}_i\) for the \(i\)-th unit is computed as:
\begin{equation}
    \mathbf{X'}_i = \frac{1}{|S_i|} \sum_{j \in S_i} \mathbf{X}_j
\end{equation}

This approach is computationally efficient, requiring only element-wise arithmetic operations. The resulting merged representation \(\mathbf{X'}\) has a shorter sequence length \(N'\), representing the number of words such as identifiers and symbols, instead of individual subword representations.

\subsubsection{Learning-based Merging}

The learning-based strategy assigns adaptive weights to subwords before merging them. This is achieved through an attention mechanism that computes a weighted average of the subword embeddings within each group. Given the token representations \(\mathbf{X}\), the attention weight \(\alpha_j\) for the \(j\)-th subword is computed as:
\begin{equation}
    \alpha_j = \frac{\exp(w^T \mathbf{X}_j)}{\sum_{k \in S_i} \exp(w^T \mathbf{X}_k)}
\end{equation}
where \(w \in \mathbb{R}^{d}\) is a learnable parameter vector. The final merged representation is obtained as a weighted sum:
\begin{equation}
    \mathbf{X'}_i = \sum_{j \in S_i} \alpha_j \mathbf{X}_j
\end{equation}

This strategy enables the model to dynamically adjust token contributions based on context. 
\subsubsection{Implementation Details}

\begin{algorithm}[t]
\caption{Vectorized Subword Grouping}
\label{alg:subword_grouping}
\small
\begin{algorithmic}[1]
\State \textbf{Input:} A batch of word ID sequences, $W_{batch}$, where $W_{batch}$ is a list of lists. Special tokens are represented by \texttt{None}.
\State \textbf{Output:} A tensor of group indices, $G$.

\Procedure{GroupSubwords}{$W_{batch}$}
    \State $\boldsymbol{W} \gets \text{Tensor}(W_{batch})$, replacing \texttt{None} with -1.
    \State $B, N \gets \text{shape}(\boldsymbol{W})$ \Comment{Batch size, Sequence length}
    
    \State $\boldsymbol{W}_{prev} \gets \text{prepend}(\boldsymbol{W}, \text{column of } -2)$ \Comment{Prepend a sentinel value}
    \State $\boldsymbol{W}_{prev} \gets \boldsymbol{W}_{prev}[:, :N]$ \Comment{Slice to match original length $N$}

    \State $\boldsymbol{I}_{new} \gets (\boldsymbol{W} \neq \boldsymbol{W}_{prev}) \lor (\boldsymbol{W} = -1)$ \Comment{Identify group boundaries}
    
    \State $\boldsymbol{C} \gets \text{cumsum}(\boldsymbol{I}_{new}, \text{dim}=1)$ \Comment{Assign incremental group IDs}
    \State $\boldsymbol{G} \gets \boldsymbol{C} - 1$ \Comment{Convert to 0-indexed group indices}
    
    \State \textbf{return} $\boldsymbol{G}$
\EndProcedure
\end{algorithmic}
\end{algorithm}

We implement the merging strategies using tokenizers offered by the \texttt{transformers} library~\cite{wolf_2020}. Specifically, we use the \texttt{word\_ids()} API that maps each subtoken to an original word. For instance, the output \texttt{[None, 1, 1, 2, None]} indicates that the subtokens at indices $1$ and $2$ (using $0$-based indexing) belong to the token at index $1$. \texttt{None} values correspond to special tokens.
A naive implementation would consist of nested for-loops that identify the groups of subtokens that represent the same token, and then apply the merging strategy on that group. However, this is inefficient and leads to an excessive computational overhead that grows as the batch size and the sequences' lengths grow.
To address this, we implement the subgrouping and merging in a vectorized form to leverage leverage GPU parallelism.

Algorithm~\ref{alg:subword_grouping} computes group indices for merging subword token representations into word-level representations in a batched setting. The input \( W_{batch} \) is a batch of word ID sequences, where each sequence is a list of integers or \texttt{None}, as generated by the \texttt{word\_ids()}. First, \( W_{batch} \) is converted to a tensor \( \boldsymbol{W} \) with \texttt{None} values replaced by $-1$. A shifted tensor \( \boldsymbol{W}_{prev} \) is then created by prepending a column of $-2$ and slicing to match the original length, aligning each token's word ID with the previous token's. Group boundaries are identified in \( \boldsymbol{I}_{new} \) where word IDs differ from the previous token or are $-1$ (special tokens). The cumulative sum of \( \boldsymbol{I}_{new} \) along the sequence dimension yields incremental group IDs, which are adjusted to start at $0$, producing the group indices tensor \( \boldsymbol{G} \). Using this output, we aggregate subword representations using the aforementioned strategies that are implemented using PyTorch's \texttt{scatter-reduce} operations. 
This vectorized implementation reduces the sequence length to one representation per word or special token in a parallelized manner.

\subsubsection{Integration with Transformer-Based Models}

\begin{algorithm}[t]
\tiny
\caption{Transformer Forward Function for Subtoken Merging}
\label{alg:transformer_forward}
\begin{algorithmic}[1]
    \Procedure{TransformerForward}{$x_{src}, x_{tgt}$, $G$}
    
    \State \common{$h_{enc} \gets \text{InputEmbedding}(x_{src}) + \text{PositionalEncoding}(x_{src})$}
    \State  \merge{$G \gets \text{GroupSubwords}(W_{batch})$}
    \If{$I$ is 0}
    \State  \merge{$memory \gets \text{Merge}(memory, G)$}
    \EndIf
    \For{\common{$l = 1 \text{ to } N$}}
        \State \common{$h_{enc} \gets \text{LayerNorm}(h_{enc} + \text{MultiHeadAttention}(Q=h_{enc}, K=h_{enc}, V=h_{enc}))$}
        \State \common{$h_{enc} \gets \text{LayerNorm}(h_{enc} + \text{FeedForward}(h_{enc}))$}
    \EndFor
    \State \common{$memory \gets h_{enc}$}
    \If{$I$ is l}
    \State  \merge{$memory \gets \text{Merge}(memory, G)$}
    \EndIf
    
    \hrulefill

    \State \encoderonly{ --- \textbf{Path 1: Encoder-Only Model Output} (e.g., for classification) ---}
    \State \encoderonly{$h_{pooled} \gets \text{Pooling}(memory)$ \Comment{Take representation of the `[CLS]` token}}
    \State \encoderonly{\textbf{return} $\text{ClassificationHead}(h_{pooled})$}
    
    \Statex
    \hrulefill
    \Statex

    \State \decoderonly{ --- \textbf{Path 2: Encoder-Decoder Model}  ---}
    \State \decoderonly{$h_{dec} \gets \text{InputEmbedding}(x_{tgt}) + \text{PositionalEncoding}(x_{tgt})$}
    
    \For{\decoderonly{$l = 1 \text{ to } N$}}
        \State \decoderonly{$h_{dec} \gets \text{LayerNorm}(h_{dec} + \text{MaskedMultiHeadAttention}(Q=h_{dec}, K=h_{dec}, V=h_{dec}))$}
        \State \decoderonly{$h_{dec} \gets \text{LayerNorm}(h_{dec} + \text{MultiHeadAttention}(Q=h_{dec}, K=memory, V=memory))$}
        \State \decoderonly{$h_{dec} \gets \text{LayerNorm}(h_{dec} + \text{FeedForward}(h_{dec}))$}
    \EndFor
    \State \decoderonly{$logits \gets \text{Linear}(h_{dec})$}
    \State \decoderonly{\textbf{return} $\text{Softmax}(logits)$}
    
    \EndProcedure
\end{algorithmic}
\end{algorithm}
Both merging strategies can be integrated into different stages of a Transformer model~(as shown in Figure~\ref{fig:method_overview}), affecting how token sequences are processed. Such integration is formalized in Algorithm~\ref{alg:transformer_forward}. It describes the steps of the \texttt{forward} pass of the Transformer model integrated with the merging layer. 

Lines highlighted in \textit{black} are the common steps performed by the encoder-only and encoder-decoder variants of the Transformer architecture. Lines in \textit{\encoderonly{blue}} represent the steps exclusively executed by the encoder-only variant, whereas those in \textit{\decoderonly{orange}} are for the encoder-decoder. Lines $4$, $7$, and $15$ in \textit{\merge{green}} are responsible for the merging-related operations discussed in Section~\ref{token_mering_strategies}. This shows how our method can be seamlessly integrated with current models with minimal effort. We experiment with two configurations related to the position of the merging operation: 

\emp{1) Post-embedding layer.} Merging is applied immediately after token embeddings are computed, before passing them into the subsequent layers.

\emp{2) Post-Transformer blocks.} Merging layers are inserted after a Transformer layer to influence intermediate representations.

\subsection{Study Subjects}

\subsubsection{Dataset Selection}

To evaluate the proposed merging strategies, we conduct experiments on three widely used code-related datasets covering different classification-based and generation-based software engineering tasks:

\emp{1) Vulnerability detection.} It is a critical SE task focused on identifying security flaws in source code that could be exploited by malicious actors. We use Fan~\etal{}~\citep{fan2020bigvul} 's \bigvul{} dataset that was curated from $348$ C/C++ projects and composed of $11,823$ vulnerable and $253,096$ non-vulnerable functions. We use a $70$:$15$:$15$ split ratio for train, validation, and test splits.

\emp{2) Code classification.} This task categorizes programs into predefined semantic classes (\eg{} algorithmic problems, such as sorting, dynamic programming). This task streamlines code organization, retrieval, and reuse in software repositories. We use the \poj{} dataset proposed by~\citep{mou_poj104_2016}. It contains C/C++ programs categorized into $104$ algorithmic problem classes.

\emp{3) Code translation.} It refers to the automated conversion of code between programming languages while preserving functionality (\eg{} Python to Java). Use cases of such a task include legacy system modernization and knowledge transfer across language ecosystems. As a result, it reduces rewrite efforts and potential errors, which subsequently lowers the cost of software migration and maintenance and enhances productivity. We use the \codetrans{} dataset from \codexglue{}~\citep{lu2021codexglue} to solve this task. It is composed of pairs of Java and C\# methods. We use the default pre-split sets of $10,300$ samples for training, $500$ for validation, and $1,000$ for testing.

\subsubsection{Model Selection}

We select the models shown in Table~\ref{tab:study_models} that represent state-of-the-art encoder-only and encoder-decoder models in solving the aforementioned tasks as of conducting this work~\cite{Zeng2022, CodeXGLUELB2021, Saad2025}. Our goal is to study post-tokenization merging, \ie{} operations that require \textit{bidirectional context} where the model can see the entire input sequence at once to know that subtokens together form a certain token. This requirement conflicts with the \textit{causal}, \textit{autoregressive} paradigm of decoder-only models. During generation, decoder-only models produce one subtoken at a time, conditioning only on the sequence generated so far. Applying our method to the generation phase would violate the causal mask, which is the defining principle of these models. For example, given a variable called ``\texttt{myVariable}'', when the model predicts the subtoken ``\texttt{my}'', it has no future context to know that the next subtoken will be ``\texttt{Variable}''. 
Because the model cannot merge tokens it has not yet produced, the very concept of merging subtokens breaks down. Hence, we apply our method during generation only in encoder-decoder models, which naturally support full-sequence context.

\begin{table}[t]
\centering
\caption{An overview of studied models.}
\label{tab:study_models}
\small
\begin{tabular}{lrr}
\toprule
Architecture & Model & \# Parameters \\ \midrule
\multirow{3}{*}{Encoder-Only} & \unx{}~\citep{guo_unixcoder_2022} & 125M \\
& \cb{}~\citep{Feng2020} & 125M \\
& \gcb{}~\citep{Guo2021} & 125M \\
\midrule
\multirow{2}{*}{Encoder-Decoder} & \ctf{}~\citep{wang2021codet5} & 220M \\ 
 & {\sc CodeT5+}~\citep{wang_codet5p_2023} & 220M, 770M \\ \bottomrule
\end{tabular}
\end{table}

\subsubsection{Evaluation Metrics}

The effectiveness of the merging strategies is assessed using two evaluation criteria: computational efficiency and task performance.

\emp{1) Computation Efficiency.} To measure the effect of token merging on the computational savings, we report the number of floating-point operations (FLOPs). This metric measures the number of arithmetic operations that are performed during a forward pass.

\emp{2) Downstream task performance.} Model performance is assessed on downstream tasks using standard evaluation metrics. For classification-based tasks we report the F1 score, whereas in generative ones, we use the CodeBLEU~\cite{Ren2020codebleu} metric.

\subsubsection{Training Details}
We implement the merging strategies by modifying the implementation of previously mentioned models provided by the \texttt{transformers v4.48.1} and \texttt{torch v2.4.1} libraries. The reported figures of floating-point operations were calculated using \texttt{fvcore v0.1.5}. To train all model variants, we have set the batch size to $16$, used a learning rate of $2e^{-5}$ and set the number of epochs to $10$. All experiments were conducted on a machine equipped with an {\sc Intel} Intel Silver $4216$ Cascade Lake CPU, an {\sc Nvidia} V$100$ with $32$GB of VRAM and $16$GB of CPU RAM. Each experiment has been repeated $3$ times using randomly selected seeds. The performance results, \ie{} F1 and CodeBLEU values, are reported as averages across the three runs.


\section{Results}\label{sec:results}
\subsection{RQ1: Computational Savings}\label{sec:rq1}

\begin{table*}[tbh]
\centering
\caption{
Comparison of the merging methods after the embedding across models and software engineering tasks. \underline{Underlined} values indicate the best-performing merging strategy. All metrics are calculated using the test sets.
}

\begin{subtable}[b]{\textwidth}
\centering
\begin{tabular}{lrrlrrlrr}
\toprule
\multirow{2.5}{*}{\begin{tabular}[c]{@{}l@{}}Merging\\ Strategy\end{tabular}} & \multicolumn{2}{c}{\cb{}} &  & \multicolumn{2}{c}{\gcb{}} &  & \multicolumn{2}{c}{\unx} \\ \cmidrule(lr){2-3} \cmidrule(lr){5-6} \cmidrule(l){8-9} 
 & F1 & FLOPs ($\times 10^{14}$) &  & F1 & FLOPs ($\times 10^{14}$) &  & F1 & FLOPs ($\times 10^{14}$) \\ \midrule
Baseline (no merging)        & 93.86 & 15.92 &  & 94.40 & 15.92 &  & 95.44 & 15.99 \\ \hdashline \addlinespace
Static merging (mean)        & 91.69 & $13.53 \, (\times 1.18)$ &  & 92.66 & $13.53 \, (\times 1.18)$ &  & \underline{95.29} & $14.99 \, (\times 1.07)$ \\
Learning-based merging       & \underline{92.04} & $13.53 \, (\times 1.18)$ &  & \underline{92.99} & $13.53 \, (\times 1.18)$ &  & 95.06 & $14.99 \, (\times 1.07)$ \\ \bottomrule
\end{tabular}
\caption{Vulnerability detection (\bigvul{})}
\end{subtable}

\vfill

\begin{subtable}[b]{\textwidth}
\centering
\begin{tabular}{lrrlrrlrr}
\toprule
\multirow{2.5}{*}{\begin{tabular}[c]{@{}l@{}}Merging\\ Strategy\end{tabular}} & \multicolumn{2}{c}{\cb{}} &  & \multicolumn{2}{c}{\gcb{}} &  & \multicolumn{2}{c}{\unx} \\ \cmidrule(lr){2-3} \cmidrule(lr){5-6} \cmidrule(l){8-9} 
 & F1 & FLOPs ($\times 10^{14}$) &  & F1 & FLOPs ($\times 10^{14}$) &  & F1 & FLOPs ($\times 10^{14}$) \\ \midrule
Baseline (no merging)        & 98.32 & 6.26 &  & 98.47 & 6.26 &  & 98.75 & 6.29 \\ \hdashline \addlinespace
Static merging (mean)        & 98.10 & $5.07 \, (\times 1.23)$ &  & 98.37 & $5.07 \, (\times 1.23)$ &  & \underline{98.74} & $5.89 \, (\times 1.07)$ \\
Learning-based merging       & \underline{98.17} & $5.07 \, (\times 1.23)$ &  & \underline{98.39} & $5.07 \, (\times 1.23)$ &  & 98.71 & $5.89 \, (\times 1.07)$ \\ \bottomrule
\end{tabular}
\caption{Code classification (\poj{})}
\end{subtable}

\vfill

\begin{subtable}[b]{\textwidth}
\centering
\begin{tabular}{lrrlrrlrr}
\toprule
\multirow{2.5}{*}{\begin{tabular}[c]{@{}l@{}}Merging\\ Strategy\end{tabular}} & \multicolumn{2}{c}{\ctf{}} &  & \multicolumn{2}{c}{\ctfpttz{}} &  & \multicolumn{2}{c}{\ctfpssz{}} \\ \cmidrule(lr){2-3} \cmidrule(lr){5-6} \cmidrule(l){8-9} 
 & CodeBLEU & FLOPs ($\times 10^{14}$) &  & CodeBLEU & FLOPs ($\times 10^{14}$) &  & CodeBLEU & FLOPs ($\times 10^{14}$) \\ \midrule
Baseline (no merging) & 68.49 & 1.281 &  & 67.41 & 1.281 &  & 67.76 & 4.149 \\  \hdashline \addlinespace
Static Merging (mean) & \underline{70.96} & $1.267 \, (\times 1.01)$ &  & 66.69 & $1.267 \, (\times 1.01)$ &  & 67.83 & $4.102 \, (\times 1.01)$ \\
Learning-based merging & 70.69 & $1.267 \, (\times 1.01)$ &  & \underline{67.67} & $1.267 \, (\times 1.01)$ &  & \underline{70.02}  &  $4.102 \, (\times 1.01)$\\ \bottomrule
\end{tabular}
\caption{Code translation Java $\rightarrow$ C\# (\codetrans{} from \codexglue{})}
\end{subtable}

\label{tab:results_emb_merging}
\end{table*}

\textbf{Token merging strategies lead to substantial computational savings across all studied model architectures and SE tasks.} As shown in Table~\ref{tab:results_emb_merging}, both static and learning-based merging strategies reduce FLOPs by up to $19$\% compared to baseline models without merging. This result is particularly promising given that no changes are required to the model architecture or tokenizer. Instead, a lightweight post-tokenization step reduces the number of tokens processed. For classification tasks such as vulnerability detection on the \bigvul{} dataset, the static merging strategy reduces FLOPs by $15$\% for \cb{} and \cb{}, and by $6.3$\% for \unx{}, compared to a no-merging baseline. Similarly, in code classification on the \poj{} dataset, savings reach $19$\% for \cb{} and \gcb{}. However, in generation tasks such as code translation on the \codetrans{} dataset, the savings are smaller, with {\sc CodeT5} models achieving a $1.1$\% reduction. Larger models, such as \ctfpssz{}, exhibit modest percentage savings but significant absolute reductions due to their high baseline costs. 
The consistent FLOPs reduction across the studied encoder-only and encoder-decoder models demonstrates the general applicability of token merging as a resource-efficient optimization. 

\begin{figure}[tbh]
    \centering
    \includegraphics[width=0.7\linewidth]{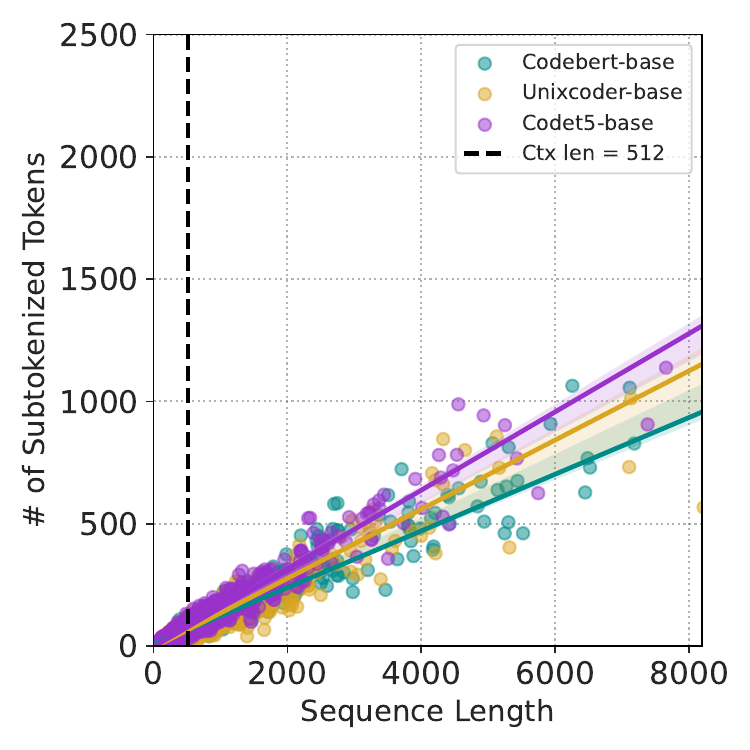}
    \caption{Regression plots of sequence lengths vs. \# of subtokenized tokens on the \bigvul{} dataset. \cb{} and \gcb{} share the same tokenizer. The same thing applies to the {\sc CodeT5X} models.}
    \label{fig:seq_len_subtokens_corr_bigvul}
\end{figure}

\textbf{Computational savings scale with sequence length due to the subtokenization overhead of BPE-based tokenizers.} Figure~\ref{fig:seq_len_subtokens_corr_bigvul} shows a strong linear relationship between the raw input sequence length and the number of subtokenized tokens across three representative models. Longer code snippets, which often include complex and compound identifiers~\cite{deissenboeck_concise_2006}, are disproportionately affected by BPE tokenization, resulting in high token inflation. Since token merging collapses subwords into a single semantic unit, it effectively compresses these longer sequences. As a result, FLOPs savings increase with the complexity and length of the input, offering even more pronounced benefits in real-world applications where longer functions and classes are common.

Current model limitations constrain the full extent of potential savings. Due to the fixed $512$-token context limit of the used pre-trained models, the experiments are bounded in terms of achievable sequence compression. However, as Transformer models increasingly support longer context windows, the merging strategies are expected to yield even greater computational reductions. Tasks such as vulnerability detection and code summarization, which frequently process extended contexts, are poised to benefit the most. These findings underscore the forward compatibility of token merging techniques with emerging trends in large-context language models.

\smrybx{\textbf{Summary}: Merging strategies significantly reduce computational requirements, with savings in FLOPs ranging from $1$\% to $19$\% depending on the task and model. The difference in savings could be attributed to a multitude of reasons, such as the nature of the dataset or the proneness of the model's tokenizer to generate a higher count of subtokens. In addition, these savings are bottlenecked by the context window of the studied models. Our approach promises greater benefits with models that support processing larger sequences.}

\subsection{RQ2: SE Task Performance}\label{sec:rq2}

\textbf{The baseline consistently delivers robust performance for the classification tasks, reflecting its ability to leverage detailed subtoken information}. On the \bigvul{} dataset, baseline F1 scores reach $93.86$ for \cb{}, $94.40$ for \gcb{}, and $95.44$ for \unx{}. The merging strategies yield \textit{slightly} lower but \textit{highly competitive} results. For example, learning-based merging records F1 scores of $92.04$ for \cb{} and $92.99$ for \gcb{}, while static merging with \unx{} reaches $95.29$, nearly identical to the baseline's $95.44$. A similar trend appears in the \poj{} task, where baseline F1 scores are $98.32$ for \cb{}, $98.47$ for \gcb{}, and $98.75$ for \unx{}. Here, learning-based merging scores $98.17$ for \cb{} and $98.39$ for \gcb{}, and static merging with \unx{} achieves $98.74$, closely approaching the baseline. These subtle differences suggest that merging strategies maintain strong performance, with the added advantage of computational savings that can enhance scalability in practical applications.

\textbf{For the generation task on the \codetrans{} dataset, merging strategies match or sometimes surpass the baseline.} For \ctf{}, static merging achieves a CodeBLEU score of $70.96$, improving upon the baseline's $68.49$, while learning-based merging scores $70.69$, also a gain over the non-merging baseline. With \ctfpttz{}, learning-based merging edges out the baseline ($67.67$ vs. $67.41$), though static merging falls slightly short. The most notable improvement occurs with \ctfpssz{}, where learning-based merging reaches a CodeBLEU of $70.02$, comfortably ahead of the baseline's $67.76$ and static merging's $67.83$. These outcomes indicate that merging strategies, particularly learning-based approaches in larger models, can enhance generative performance, possibly by fostering more cohesive token representations in the output.

These results reveal nuanced insights. In classification tasks, where fine-grained distinctions are key, the baseline's preservation of subtoken details offers a \textit{slight} edge, yet merging strategies remain \textit{remarkably close} in performance while reducing computational overhead. In generation tasks, the flexibility of learning-based merging appears to benefit larger models such as \ctfpssz{}, suggesting that trainable aggregation can optimize output quality. Model-specific behaviors also emerge: \unx{}, for instance, adapts well to static merging in classification, likely due to its architectural strengths, while {\sc codeT5} variants perform better with learning-based merging in generation tasks.
\smrybx{\textbf{Summary}: While the baseline often excels in classification tasks, merging strategies provide a compelling alternative, delivering near-comparable performance alongside significant computational benefits. In generation tasks, these strategies can even outperform the baseline. These findings highlight a practical trade-off: the minor performance differences are more than offset by efficiency gains, making merging strategies an attractive choice for optimizing resource use without substantially compromising effectiveness.}

\subsection{RQ3: Impact of Layer Position}\label{sec:rq3}

\begin{figure*}
    \centering
    \begin{subfigure}[t]{0.33\textwidth}
        \includegraphics[width=\linewidth]{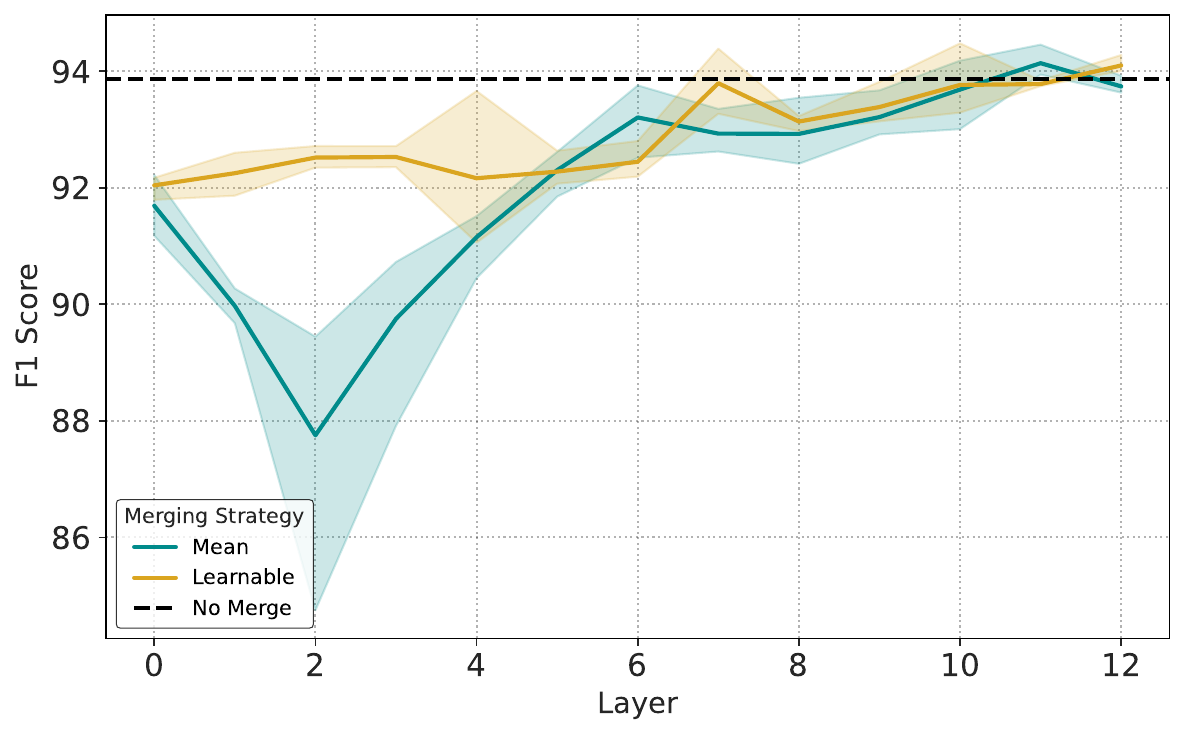}
        \caption{\cb{}}
        \label{fig:cb_bigvul}
    \end{subfigure}
    \hfill
    \begin{subfigure}[t]{0.33\textwidth}
        \includegraphics[width=\linewidth]{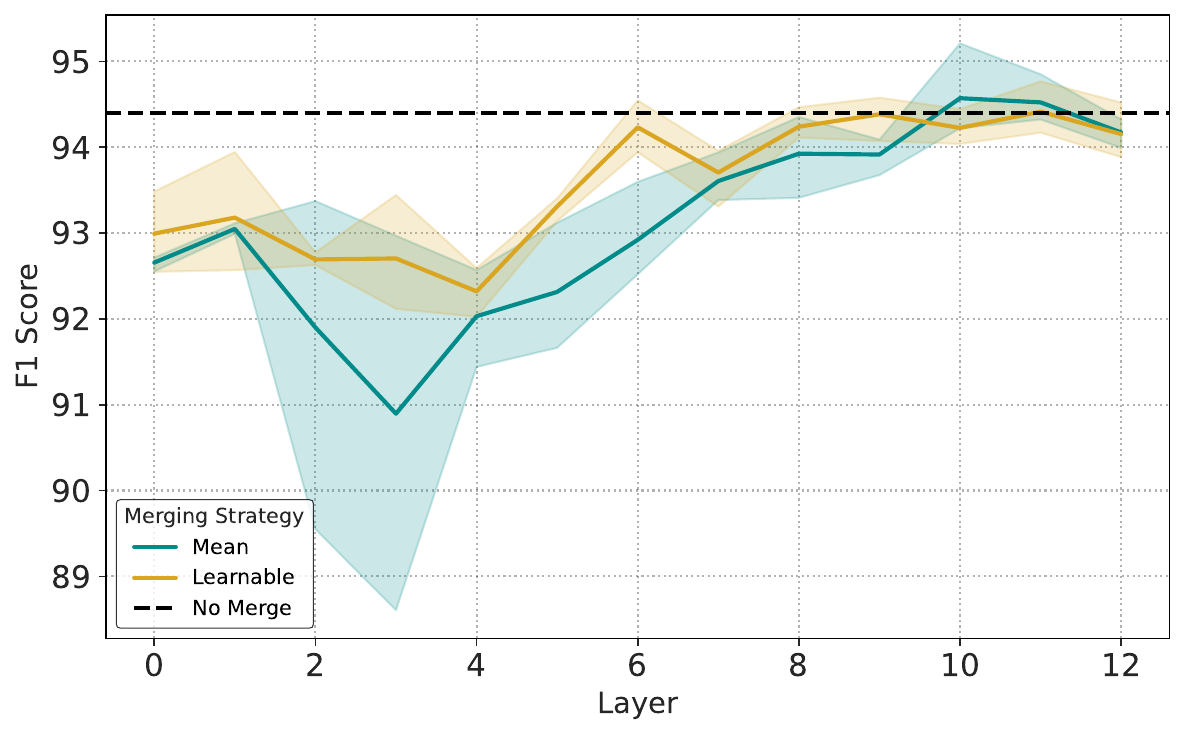}
        \caption{\gcb{}}
        \label{fig:gcb_bigvul}
    \end{subfigure}
    \hfill
    \begin{subfigure}[t]{0.33\textwidth}
        \includegraphics[width=\linewidth]{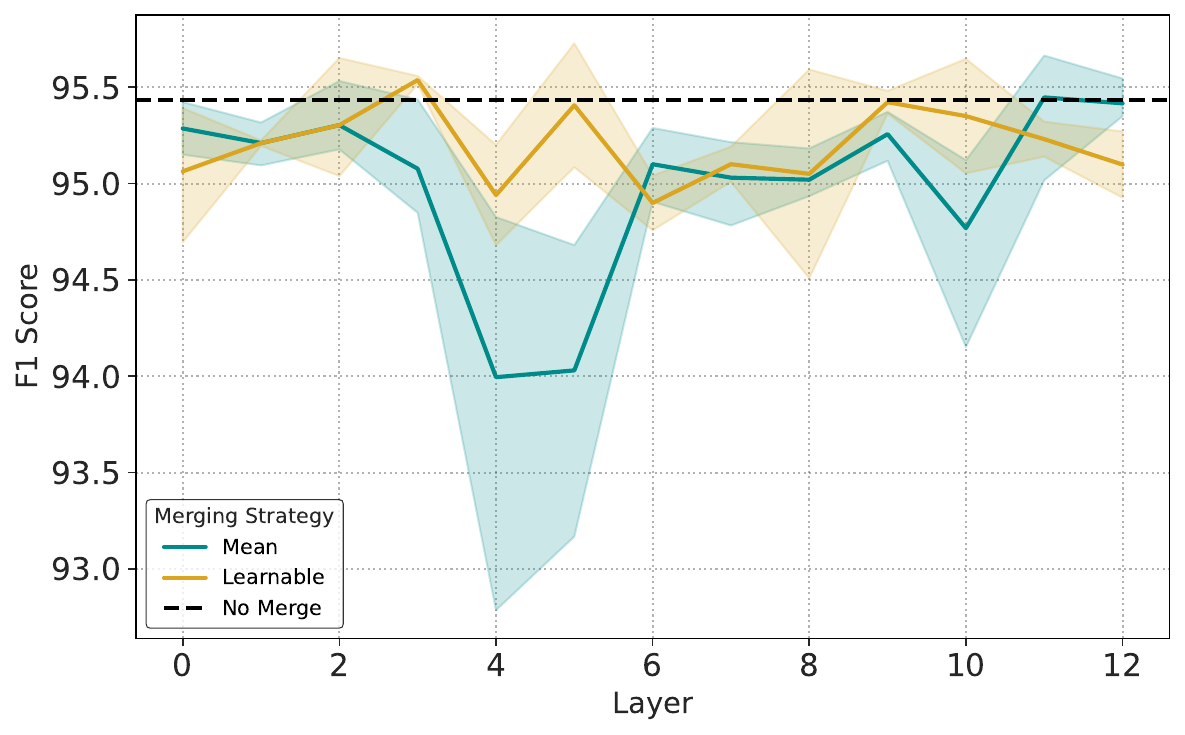}
        \caption{\unx{}}
        \label{fig:unx_bigvul}
    \end{subfigure}
    \caption{Performance of \cb{}, \gcb{} and \unx{} on the \bigvul{} dataset across all merging strategies and layers.}
    \label{fig:bigvul_results}
\end{figure*}
\begin{figure*}
    \centering
    \begin{subfigure}[t]{0.33\textwidth}
        \includegraphics[width=\linewidth]{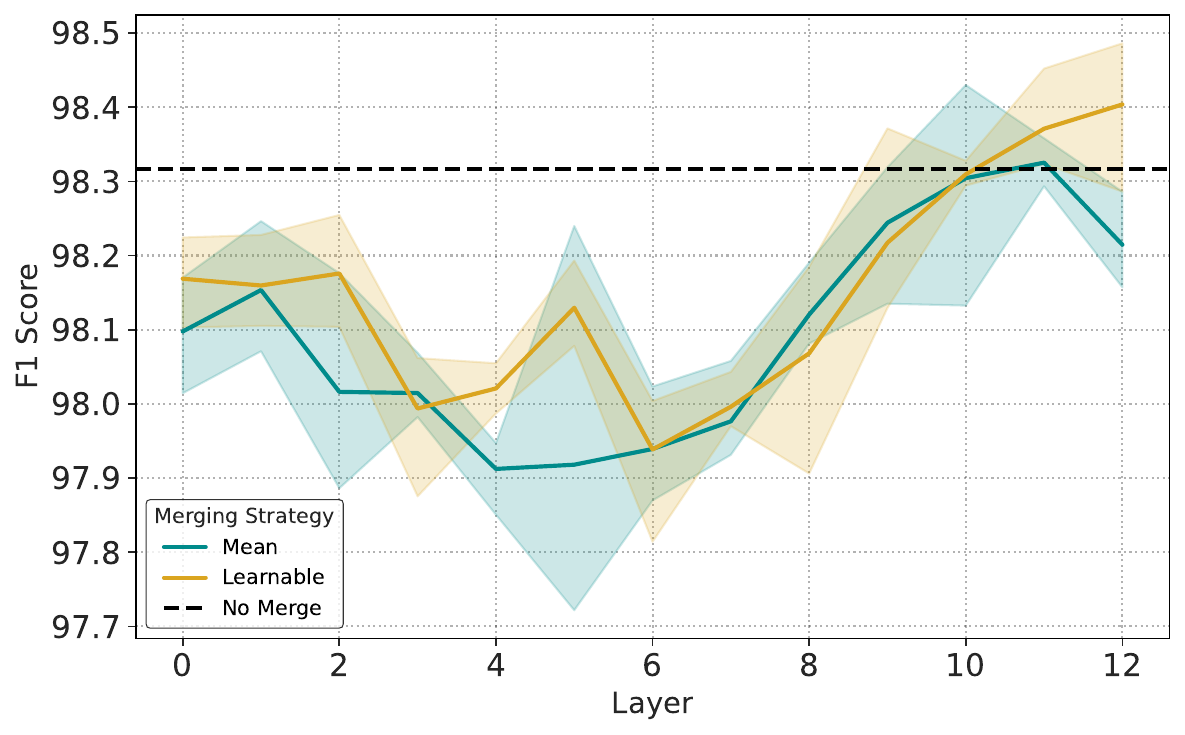}
        \caption{\cb{}}
        \label{fig:cb_poj}
    \end{subfigure}
    \hfill
    \begin{subfigure}[t]{0.33\textwidth}
        \includegraphics[width=\linewidth]{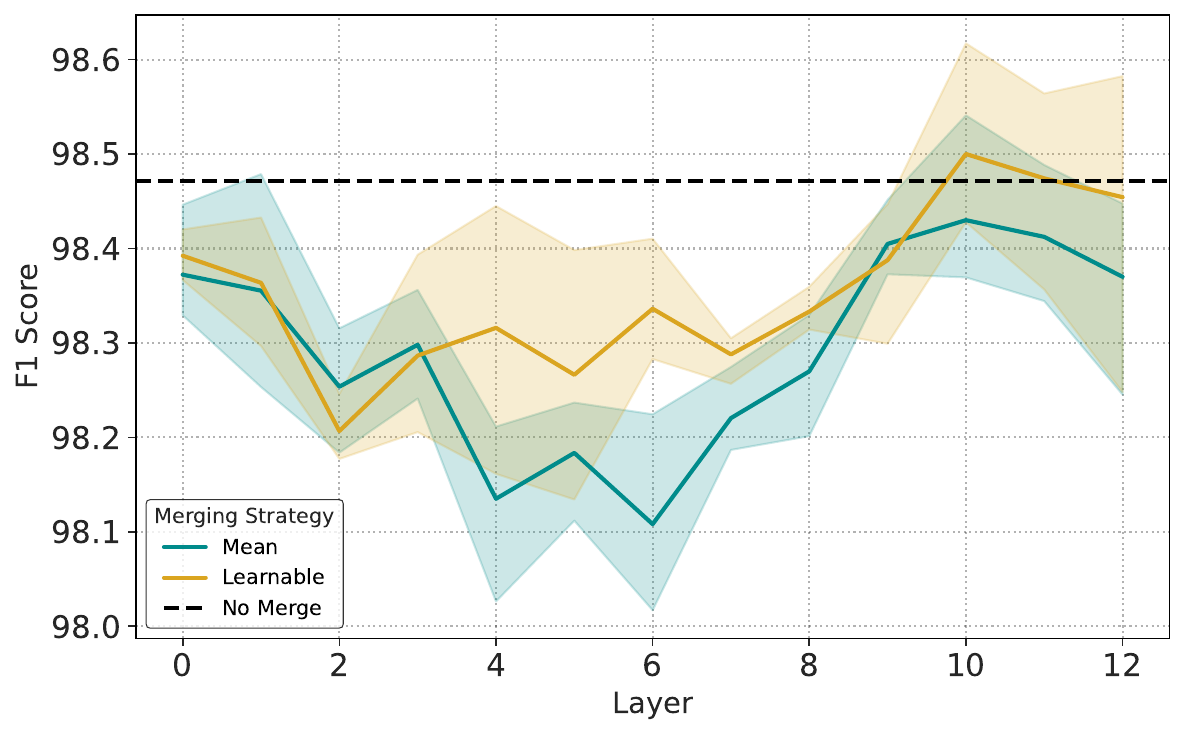}
        \caption{\gcb{}}
        \label{fig:gcb_poj}
    \end{subfigure}
    \hfill
    \begin{subfigure}[t]{0.33\textwidth}
        \includegraphics[width=\linewidth]{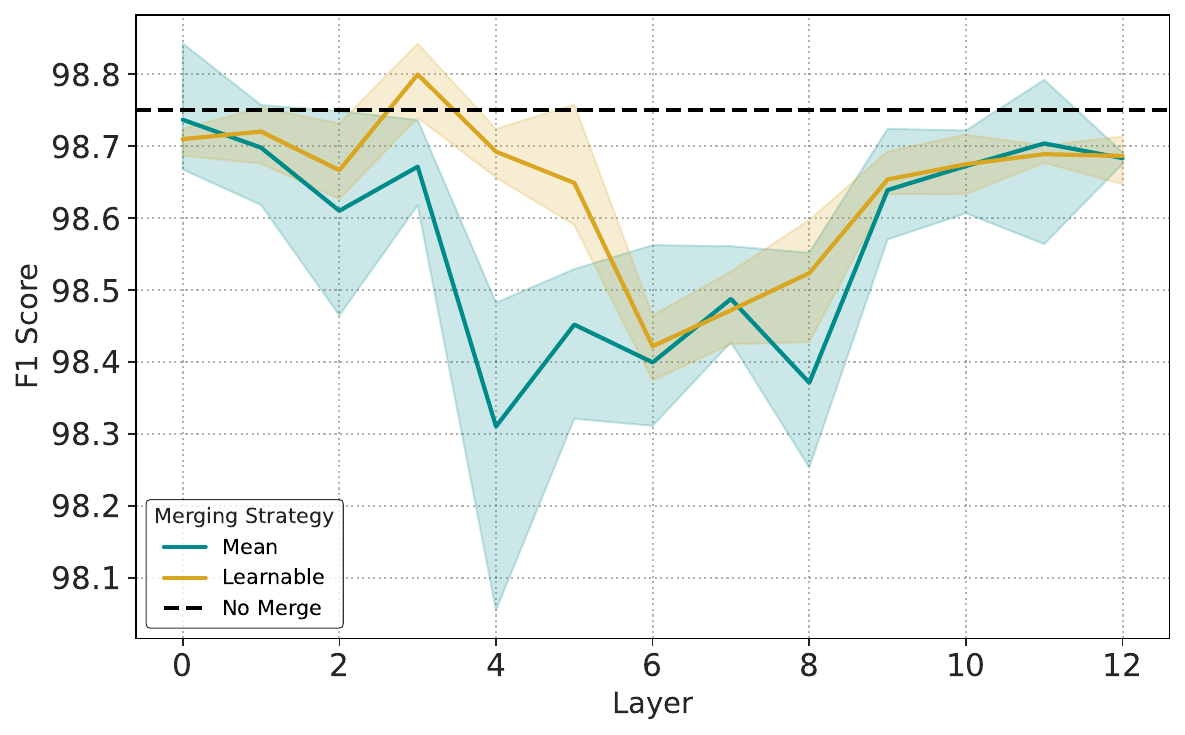}
        \caption{\unx{}}
        \label{fig:unx_poj}
    \end{subfigure}
    \caption{Performance of \cb{}, \gcb{} and \unx{} on the \poj{} dataset across all merging strategies and layers.}
    \label{fig:poj_results}
\end{figure*}
\begin{figure*}
    \centering
    \begin{subfigure}[t]{0.33\textwidth}
        \includegraphics[width=\linewidth]{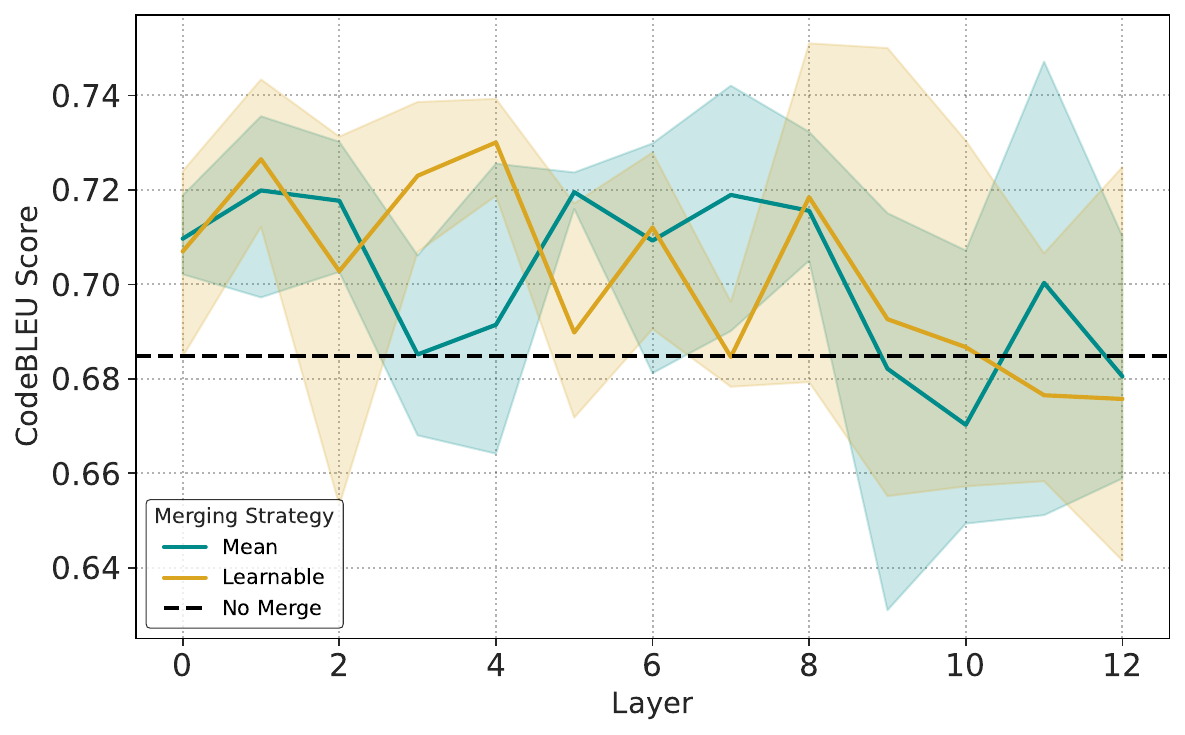}
        \caption{\ctf{}}
        \label{fig:ctf_ct}
    \end{subfigure}
    \hfill
    \begin{subfigure}[t]{0.33\textwidth}
        \includegraphics[width=\linewidth]{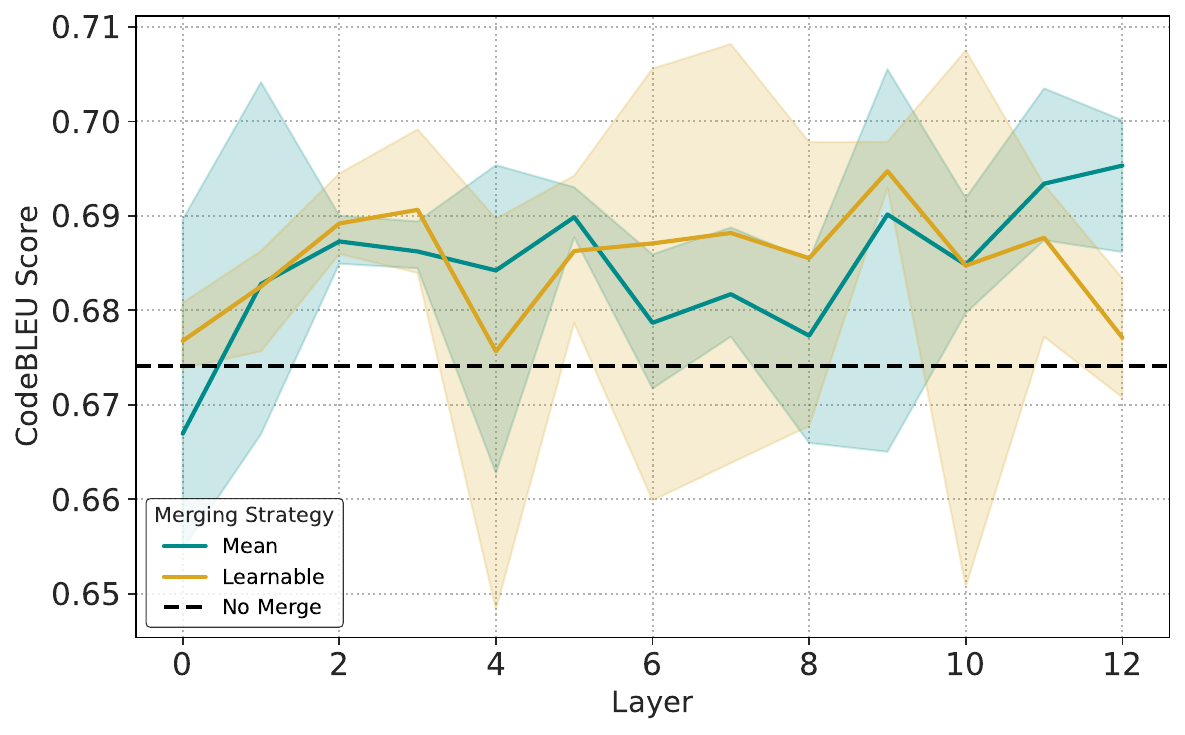}
        \caption{\ctfpttz{}}
        \label{fig:ctfp220_ct}
    \end{subfigure}
    \hfill
    \begin{subfigure}[t]{0.33\textwidth}
        \includegraphics[width=\linewidth]{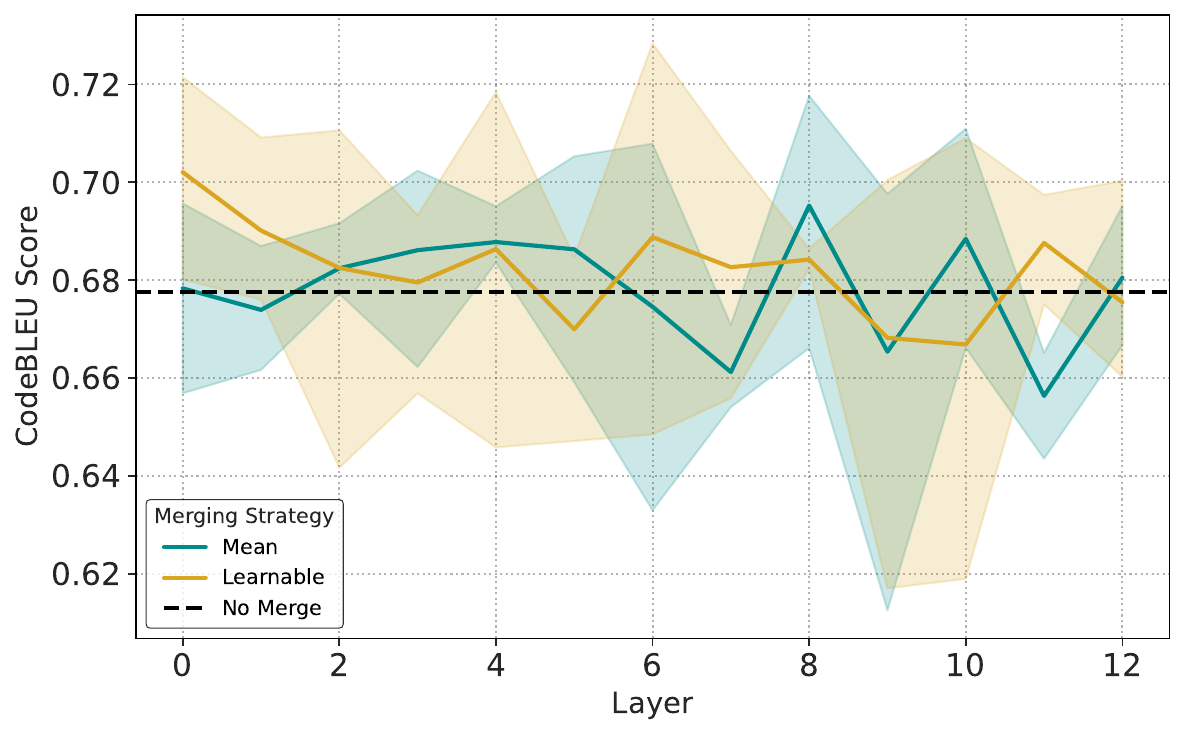}
        \caption{\ctfpssz{}}
        \label{fig:ctfp770_ct}
    \end{subfigure}
    \caption{Performance of \ctf{}, \ctfpttz{} and \ctfpssz{} on the \codetrans{} dataset across all merging strategies and layers.}
    \label{fig:codetrans_results}
\end{figure*}

Figures~\ref{fig:bigvul_results},~\ref{fig:poj_results}  and \ref{fig:codetrans_results} illustrate the performance of each merging strategy depending on its position within the language model. In all figures, the $0^{\text{th}}$ layer refers to the embedding layer.

\textbf{Across both the \bigvul{} and \poj{} datasets, a notable trend emerges concerning the placement of the merging layer.} When merging occurs early in the Transformer stack, specifically between the embedding layer and L$4$, performance typically falls below the baseline, with the static mean strategy showing particularly poor results. This suggests that combining subtokens too soon, before the model has fully developed contextualized representations, discards valuable subtoken-level details essential for understanding code. Conversely, merging in later layers, such as L$10$ to L$12$, often yields performance that matches or surpasses the baseline. By this stage, the model has constructed rich, contextual embeddings, allowing merging to consolidate information effectively without significant loss. Furthermore, the learnable strategy consistently outperforms static mean merging, especially in earlier layers, likely because its adaptability enables it to better tailor the combination of subtokens to the specific task and dataset.

Regarding the performance of \cb{}, a clear preference for late merging emerges across both datasets. On \bigvul{}, the learnable strategy begins below the baseline at the embedding layer but improves progressively, overtaking the baseline by L$12$. A parallel trend appears on \poj{}, where learnable merging also starts slightly under the baseline but peaks at L$12$, exceeding it. The mean strategy, while improving in later layers, rarely surpasses the baseline consistently. This indicates that \cb{} thrives when it can first build detailed contextual representations of code before merging.

\gcb{} follows a similar trajectory, with late merging proving advantageous. On the \bigvul{} dataset, an interesting deviation occurs: the mean merging slightly outperforms the baseline at L$10$ and L$11$, suggesting that for this model, averaging highly contextualized representations can be surprisingly effective. On \poj{}, the learnable strategy edges above the baseline at layer L$10$, while mean merging remains slightly below it. 


\unx{}, however, stands out with its stability across merging layers. On \bigvul{}, both merging strategies maintain performance close to the baseline, with learnable merging peaking early at L$3$ and slightly exceeding it. On \poj{}, learnable merging again performs well at L$3$, surpassing the baseline and holding strong thereafter. This early effectiveness hints that \unx{} architecture or pretraining equips it to handle merged representations efficiently even in initial layers, reducing its reliance on late merging compared to the other models.

\textbf{On the \codetrans{} dataset, merging strategies demonstrate robustness across all layers, frequently outperforming the no-merge baseline.} Across the \ctf{}, \ctfpttz{}, and \ctfpssz{} variants, merging generally maintains or surpasses baseline performance, in contrast to the early-layer degradation observed on \bigvul{} and \poj{}.

For \ctf{}, learnable merging achieves CodeBLEU scores above $0.73$, with performance peaking at deeper layers (e.g., L$8$ and L$9$). Notably, performance gains are also evident in early layers (L$1$ and L$3$), suggesting that judicious early merging can provide semantic compression without hindering contextual modelling.

In \ctfpttz{}, both merging strategies outperform the no-merge baseline across most layers. Learnable merging shows slight superiority, particularly from L$6$ to L$11$. Compared to \ctf{}, \ctfpttz{} shows lower variance while having the same number of parameters, indicating enhanced stability and better integration of merged representations.
The large \ctfpssz{} model demonstrates remarkable stability across all layers. Both mean and learnable strategies perform comparably, often matching or exceeding the baseline. While learnable merging marginally outperforms the baseline in several early-to-mid layers~(e.g., L$0$ to L$7$), mean merging also performs competitively, indicating that even static averaging can be effective at scale.

The tasks themselves also influence the merging dynamics. Vulnerability detection (\bigvul{}), which demands pinpointing subtle code flaws, benefits from delaying merging until later layers, where the model has fully grasped the code's contextual nuances. Code classification on \poj{}, focused on discerning high-level functional patterns, similarly favors late merging to leverage comprehensive representations. Yet, \unx{} ability to perform well with early merging on both tasks suggests that certain model designs can mitigate task-specific demands on layer placement, offering flexibility in application. In contrast, code translation~(\codetrans{}) requires capturing syntactic structure and maintaining semantic equivalence across languages. All merging strategies can be applied earlier or throughout the model to outperform or remain on par with the no-merge baseline, underscoring the potential of token merging for efficient input compression without sacrificing generative quality.

\smrybx{\textbf{Summary}: Optimal performance requires tailoring the merging strategy to the task: classification benefits from late merging (layers $10$-$12$), while code translation is more flexible and improves with early merging. Across all tasks, the learnable strategy consistently outperforms simple averaging, proving particularly effective for translation at any layer.
}

\subsection{RQ4: Performance-Computation Trade-off}\label{sec:rq4}



To select the optimal placement for our subtoken merging layer, we address a multi-objective optimization problem where we aim to simultaneously maximize model performance while minimizing computational cost. We formalize this selection process using the concept of \textit{Pareto efficiency}~\cite{NBERw20883}.

Let $S$ denote the set of all possible configurations, where each configuration $s\in S$ corresponds to placing the merging layer at a different position in the model architecture. For each configuration $s$, we measure its computational cost, $C(s)$, quantified by the number of FLOPs, and its performance, $P(s)$, quantified by the F1 or CodeBLEU score. The objective is to find a configuration $s^{*}$ that represents the best trade-off between these two competing goals.

We define the optimal trade-offs using the concept of \textit{Pareto dominance}~\cite{Li2023multi}. A configuration $s_1 \in S$ is said to Pareto-dominate another configuration $s_2 \in S$ if $s_1$ is as good as $s_2$ in all objectives and strictly better in at least one. Formally, $s_1$ dominates $s_2$ if: 
$$(P(s_1) \geq P(s_2) \land C(s_1) < C(s_2)) \lor (P(s_1) > P(s_2) \land C(s_1) \leq C(s_2))$$

A configuration $s^{*} \in S$ is considered \textit{Pareto-optimal} if no other configuration in $S$ dominates it. The set of all such Pareto-optimal points is known as the \textit{efficiency frontier}. This frontier represents the set of all configurations for which an improvement in one objective can only be achieved by degrading the other objective. Any configuration not on this frontier is suboptimal, as there exists at least one other configuration that provides better performance for the same or lower cost. Given our finite set of evaluated configurations, we identify the efficiency frontier using Algorithm~\ref{alg:effiency_frontier}.

\begin{algorithm}[t]
\caption{Algorithm for Identifying the Efficiency Frontier}
\label{alg:effiency_frontier}
\begin{algorithmic}[1]
\State \textbf{Input} A set $S_{\text{eval}} = \{(c_i, p_i)\}_{i=1}^n$ of evaluated configurations, where $c_i$ is the cost and $p_i$ is the performance of configuration $i$.
\State Sort $S_{\text{eval}}$ by cost $c_i$ in ascending order, and then by performance $p_i$ in descending order to obtain the list $S'_{\text{eval}}$.
\State Initialize an empty set $S^*$
\State Initialize $p_{\max} = -\infty$
\For{each $(c, p)$ in $S'_{\text{eval}}$}
\If{$p > p_{\max}$}
\State Add $(c, p)$ to $S^*$
\State Set $p_{\max} = p$
\EndIf
\EndFor
\State The efficiency frontier is $S^*$
\end{algorithmic}
\end{algorithm}
While all points on the efficiency frontier $S^{*}$ are optimal in a Pareto sense, a single configuration must be chosen for the final model. We use a heuristic based on the \textit{knee point} of the frontier to do so. This point represents the region of diminishing returns, where the marginal gain in performance per unit of additional cost begins to decrease substantially. We identify the knee point as the configuration on the frontier that maximizes the orthogonal distance to the line connecting the two extreme points of the frontier. Let $s_{min} = (c_{min}, p_{min})$ be the frontier point with the minimum cost, and let $s_{max} = (c_{max}, p_{max})$ be the point with the maximum performance. The line $L$ connecting these two points is given by:
\[
(p_{max} - p_{min})c - (c_{max} - c_{min})p + c_{max}p_{min} - c_{min}p_{max} = 0
\]
For each point $s_i = (c_i, p_i) \in S^{*}$, its orthogonal distance $d_i$ to the line $L$ is calculated as:
\[ d_i = \frac{ \left| (p_{\mathrm{max}} - p_{\mathrm{min}}) c_i - (c_{\mathrm{max}} - c_{\mathrm{min}}) p_i + c_{\mathrm{max}} p_{\mathrm{min}} - c_{\mathrm{min}} p_{\mathrm{max}} \right| }{ \sqrt{ (p_{\mathrm{max}} - p_{\mathrm{min}})^2 + (c_{\mathrm{max}} - c_{\mathrm{min}})^2 } } \]
The configuration $s_{knee}$ selected as the optimal point is the one that maximizes this distance: 
\[
s_{knee} = argmax_{s_i \in S^{*}} d_i
\]

\begin{figure}[t]
        \includegraphics[width=.75\columnwidth]{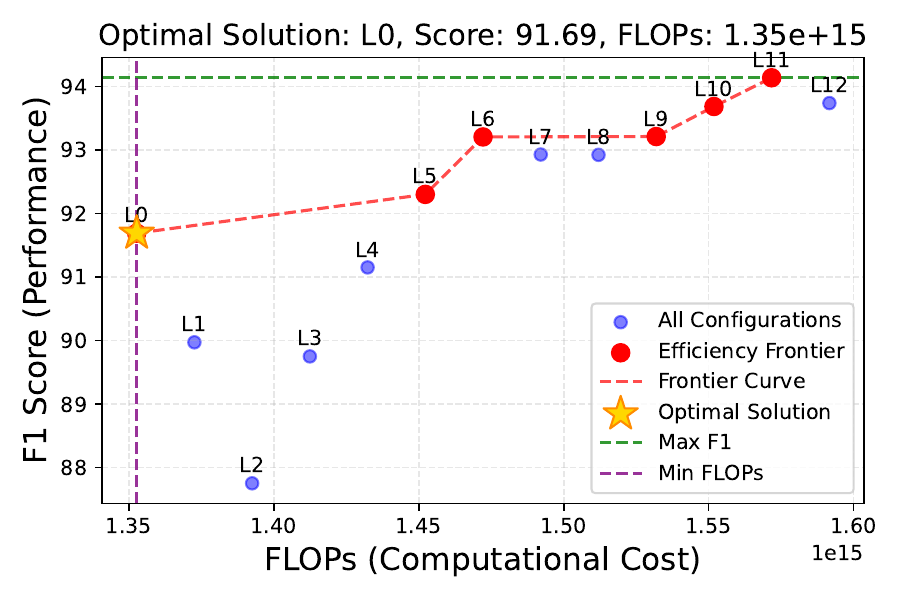}
    \caption{Example of Pareto a front of the \cb{} (mean) configuration on the \bigvul{} dataset.}
    \label{fig:frontier_bigvul}
\end{figure}

Given space limitations, we present only one example of the generated Pareto frontier plots, shown in Figure~\ref{fig:frontier_bigvul}. We refer the reader to our replication package for the rest of the figures. In addition, Table~\ref{tab:table_of_tradeoffs} summarizes the best position where merging should occur for each merging strategy across all models and tasks.

\begin{table}[b]
\centering
\caption{Summary of configurations that yield the best trade off between performance and computational savings.}
\label{tab:table_of_tradeoffs}
\resizebox{\columnwidth}{!}{%
\begin{tabular}{lllccc}
\toprule
\textbf{Task} & \textbf{Strategy} & \textbf{Model} & \textbf{Position} & \makecell{\textbf{Performance} \\ \textbf{ (F1/CodeBLEU)}} & \makecell{\textbf{FLOPs} \\ ($\times10^{14}$)} \\
\midrule
\multirow{6}{*}{\bigvul{}} 
    & Mean & \cb{} & $0$ & $91.69$ & $13.5$ \\
    & Learnable & \cb{} & $7$ & $93.79$ & $14.9$ \\
    \cmidrule(lr){2-6}
    & Mean & \gcb{} & $1$ & $93.05$ & $13.7$ \\
    & Learnable & \gcb{} & $6$ & $94.23$ & $14.7$ \\
    \cmidrule(lr){2-6}
    & Mean & \unx{} & $0$ & $95.29$ & $15.0$ \\
    & Learnable & \unx{} & $3$ & $95.54$ & $15.2$ \\
\midrule
\multirow{6}{*}{\poj{}}
    & Mean & \cb{} & $1$ & $98.15$ & $5.17$ \\
    & Learnable & \cb{} & $0$ & $98.17$ & $5.07$ \\
    \cmidrule(lr){2-6}
    & Mean & \gcb{} & $0$ & $98.37$ & $5.07$ \\
    & Learnable & \gcb{} & $0$ & $98.38$ & $5.07$ \\
    \cmidrule(lr){2-6}
    & Mean & \unx{} & $0$ & $98.74$ & $5.89$ \\
    & Learnable & \unx{} & $1$ & $98.72$ & $5.92$ \\
\midrule
\multirow{6}{*}{\codetrans{}}
    & Mean & \ctf{} & $1$ & $71.99$ & $1.27$ \\
    & Learnable & \ctf{} & $1$ & $72.65$ & $1.27$ \\
    \cmidrule(lr){2-6}
    & Mean & \ctfpttz{} & $2$ & $68.73$ & $1.27$ \\
    & Learnable & \ctfpttz{} & $3$ & $69.06$ & $1.27$ \\
    \cmidrule(lr){2-6}
    & Mean & \ctfpssz{} & $3$ & $68.61$ & $4.11$ \\
    & Learnable & \ctfpssz{} & $0$ & $70.21$ & $4.10$ \\
\bottomrule
\end{tabular}%
}
\end{table}
The results from the Pareto analysis reveal a consistent pattern: the merging layer achieves the best trade-off when positioned early in the model, typically after the embedding layer (L$0$) or one of the first few Transformer layers (L$1$-L$3$). This trend holds across different models and for both classification and translation tasks. The learnable attention-based merging strategy consistently yields better performance than static averaging, often with negligible increase in FLOPs. The most notable exception is the \bigvul{} task, where the learnable strategy finds its best trade-off much later in the network (L$6$ and L$7$), suggesting that complex vulnerability detection benefits from merging more contextualized representations.

\smrybx{\textbf{Summary:} For the best performance-cost trade-off, subtoken representations should be merged early in the network, with a learnable attention-based strategy generally outperforming simple averaging.}
\section{Discussion and Implications}\label{sec:implications}
\vspace{1mm}
\textbf{Token merging offers a drop-in approach to save computational resources.}  
For practitioners deploying language models for code in production environments under resource constraints or with latency requirements, token merging provides an immediate, low-risk way to reduce costs~(Section~\ref{sec:rq1}). This strategy avoids retraining, architecture changes, or tokenizer modifications, allowing seamless integration with existing inference pipelines. Given the quadratic scaling of attention mechanisms~\cite{vaswani2017attention}, a $30$\% reduction in token count can nearly halve attention-related FLOPs, which is especially significant for models with hundreds of millions of parameters. This provides an effective means to reduce inference costs and the associated carbon footprint in real-world deployments.

\medskip
\textbf{Learning-based merging opens avenues for task-adaptive compression without compromising generation quality in encoder-decoder language models.} Unlike static merging, which applies a uniform reduction rule, learning-based merging adjusts token salience dynamically, offering task-aware and identifier-aware condensation of token representations~(Section~\ref{sec:rq2}). Researchers can explore integrating learning-based merging with other training objectives~(e.g., contrastive learning) to evaluate its effectiveness. In addition, researchers can investigate the application of merging strategies in low-resource programming languages~\cite{cassano2024lowresourcepl}, such as Lua and R.

\medskip
\textbf{Merging strategies can be modularized and combined with existing adaptation techniques.}
While static merging reduces sequence length, Low-Rank Adaptation (LoRA)~\cite{hu2022lora} enables model adaptation with minimal additional parameters. Practitioners can use static merging together with LoRA to train models that are both efficient in processing and lightweight in fine-tuning. In the case of learning-based merging, the merging weights can be incorporated into LoRA adapters, preserving the frozen backbone. This two-tier optimization approach supports modular, resource-aware deployments, making it particularly suitable for real-world applications with strict memory and latency requirements.


\medskip
\textbf{For better performance, layer placement should follow task granularity: merge late for classification, merge early for generation.} The effect of merging layer placement highlights how different tasks interact with token granularity~(Section~\ref{sec:rq3} and Section~\ref{sec:rq4}). Practitioners can treat the merging layer position as a tunable hyperparameter that should be co-optimized with task formulation and model architecture. We suggest that for tasks requiring fine-grained syntactic sensitivity, such as vulnerability detection or code classification, deferring merging to deeper layers preserves token-level distinctions critical for accuracy. For generation tasks, early merging does not degrade performance and may even enhance output quality by promoting holistic sequence representations.


\section{Threats to Validity}\label{sec:threats}
In this section, we elaborate on the different threats to the validity of this study and the procedures we took to reduce such threats.

\medskip
\noindent\textbf{\textit{Internal Validity.}}
Internal validity refers to the degree to which the experimental design and execution ensure that the observed effects are caused by the manipulated variables rather than extraneous factors. In this study, two key threats to internal validity emerge. The first is related to the \textit{potential} implementation errors in merging strategies. To avoid such a threat, authors discussed the implementation provided by each one in the form of code reviews through pull requests. In addition, unit tests were used to verify the correctness of the output of each procedure using manually verified input and output. Moreover, we relied to the fullest extent on available heavily-tested APIs exposed by the libraries mentioned in the study. The second threat is related to the choice of hyperparameters. We used the same values from similar studies~\cite{lu2021codexglue} across all experiments. The same applies to the training and test splits. Finally, we repeated our experiments three times to mitigate bias related to randomness.

\medskip
\noindent\textbf{\textit{External Validity.}}
External validity relates to the generalizability of the study's findings beyond the specific context tested. To mitigate this, we have considered tasks that are classification-based and generative that involve three programming languages (C++, C\#, and Java), across two variants of the Transformer architecture, encoder-only and encoder-decoder.

\medskip
\noindent\textbf{\textit{Construct Validity.}}
Construct validity assesses whether the study measures what it intends to measure. We used well-established metrics to measure performance depending on the task, \ie{} F1 and CodeBLEU~\cite{Ren2020codebleu}, and FLOPs count to measure computational efficiency~\cite{Saad2025}.

\section{Conclusions}\label{sec:conclusion}
In this study, we tackled the computational inefficiency of language models for code stemming from Byte Pair Encoding tokenization, which generates lengthy subtoken sequences. Through two proposed token merging strategies, averaging and learning-based we achieved up to a $19$\% reduction in FLOPs count compared to baseline models, while largely preserving or even enhancing performance, notably in code translation with a $2.47$-point CodeBLEU increase. These findings validate token merging as an effective solution to mitigate BPE inefficiencies, with practical implications for deploying models in resource-limited settings like consumer-grade GPUs. However, the study's scope was confined to specific tasks and models. Future work could extend these strategies to diverse architectures or refine merging techniques for greater performance and interpretability. Ultimately, token merging offers a compelling balance of efficiency and effectiveness, advancing the scalability of language models in software engineering applications.

\bibliographystyle{ACM-Reference-Format}
\bibliography{main}

\end{document}